\documentclass[submitting]{nst}

\usepackage{subfigure,dcolumn}
\usepackage[T2A,T1]{fontenc}
\usepackage[russian,english]{babel}
\usepackage{graphicx}

\usepackage{listings}
\begin{document}

\title{The Cross Section Calculation of the $^{112}$Sn($\alpha$,$\gamma$)$^{116}$Te Reaction with Different Nuclear Models at the Astrophysical Energy Range}

\author{C. Yal\c{c}\i{n}}
\email[Corresponding author, ]{caner.yalcin@kocaeli.edu.tr}
\affiliation{Kocaeli University, Department of Physics, Umuttepe 41380, Kocaeli, Turkey}

\begin{abstract}
The theoretical cross section calculations for the astrophysical $p$ process are needed because the most of the related reactions are technically very difficult to be measured in the laboratory. Even if the reaction was measured, most of the measured reactions have been carried out at the higher energy range from the astrophysical energies. Therefore, almost all cross sections needed for $p$ process simulation has to be theoretically calculated or extrapolated to the astrophysical energies. The $^{112}$Sn($\alpha$,$\gamma$)$^{116}$Te is an important reaction for the $p$ process nucleosynthesis. The theoretical cross section of the $^{112}$Sn($\alpha$,$\gamma$)$^{116}$Te reaction was investigated for different global optical model potentials, level density and strength function models at the astrophysically interested energies. Astrophysical $S$ factors were calculated and compared with experimental data available in EXFOR database. The calculation with the optical model potential of the dispersive model by Demetriou et al., and Back-shifted Fermi gas level density model and Brink-Axel Lorentzian strength function model best served to reproduce experimental results at astrophysically relevant energy region. The reaction rates were calculated with these model parameters at the $p$ process temperature and compared with the current version of the reaction rate library Reaclib and Starlib.
\end{abstract}


\maketitle

 \nolinenumbers

\section{Introduction}
\label{intro}

Although there have been significant studies conducted to explain nucleosynthesis questions still remain on production of the nuclei heavier than iron. The $p$  process (or $\gamma$-process) is one of the astrophysical processes that is responsible for the production of proton rich nuclei. These nuclei are located along the proton-rich side of the stability line between Se and Hg and these nuclei are referenced as p-nuclei \cite{Woosley78, Arnould03, Rauscher13}. Burbidge et al. \cite{Burbidge57} and Cameron \cite{Cameron57} suggested that $p$ nuclei are produced  by the massive stars through photodecomposition at very high temperature in stellar environment. The production mechanism is composed of mostly ($\gamma$,n), ($\gamma$,p) and ($\gamma,\alpha$) reactions on preexisting $s$ and $r$ seed nuclei in the temperature range between \mbox{2 GK} and \mbox{3 GK}. \cite{Woosley78,Rapp06,Rauscher06,Mohr07}. The astrophysically relevant energy range for the charged particle induced nuclear reactions is called  Gamow window. The Gamow windows of astrophysical reactions were numerically calculated by Rauscher \cite{Rauscher10}. 

In order to simulate the $p$ process, it is required that a large set of information be known, including nuclear parameters. This information consists of the accurate initial seed abundances which are coming from $s$ and $r$-process model calculations, the description of the stellar medium, and the nuclear properties such as reaction cross sections, reaction rates, nuclear masses, and decay rates. This information is needed directly or indirectly for the $p$ process network simulation. In view of the nuclear parameters, the reaction rates which are derived from cross sections of more than 20,000 reactions involving about 2000 nuclei are needed for the $p$ process simulation \cite{Sauter97}. However, there are few reaction cross sections experimentally measured at the astrophysical relevant energies. The reason for this is most of the related reactions need a radioactive target and beam, also the cross sections at the astrophysical energies are very small to measure with current technology. Because of these experimental limitations, almost all reaction cross sections (or the reaction rates) needed for the $p$ process must be calculated theoretically.

On the other hand, measured experimental cross sections are needed to be extrapolated to the astrophysical energies, because most of the experiments have been performed at higher energies than astrophysical energies \cite{Gyurky2010, Scholz2014, Yalcin2015, Guray2015, Halasz2016, Mayer2016}. An alternative method to the extrapolation of the experimental cross sections to lower energies is to calculate the cross sections theoretically by using the best nuclear parameters that are deduced from the comparison of the experimental results at energies close to the astrophysical energies. Then, using the best parameters, the cross sections can be calculated for the all energy ranges relevant to the $p$ process.    

The Sn-112 is an important p-nucleus and the cross section measurements were experimentally performed at energies close to astrophysically relevant energies using different methods \cite{Ozkan02, Ozkan07, Rapp08, Netterdon15}. The Gamow window of the $^{112}$Sn($\alpha$,$\gamma$)$^{116}$Te reaction is between $\mbox{$E_{Lab.}$ = 6.38}$ and $\mbox{$E_{Lab.}$ = 10.07 $MeV$}$ at the temperature of $\mbox{3 $GK$}$. The $^{112}$Sn has also special importance because it has a magic proton number (Z=50) and it is a closed-shell nucleus. 
Consequently, the $^{112}$Sn($\alpha$,$\gamma$)$^{116}$Te reaction was chosen for investigation in order to understand the effect of different nuclear models entering the cross section calculations, such as nucleon-nucleus optical model potentials (OMP), level density models (LDM) and $\gamma$-ray strength function models (SFM).

The main steps of this study are (1) investigation of the global nuclear models effects which are entering the cross section calculation, (2) comparison with the experimental data and suggesting best global parameters, and (3) calculating the reaction rate with the best model parameters of the $^{112}$Sn($\alpha$,$\gamma$)$^{116}$Te reaction and comparing with currently used reaction rates of the Reaclib $v2.2$ \cite{Reaclib} and Starlib $v6$ \cite{Starlib}. 

\section{Model calculations and Results}
\label{calc}

The nuclear model calculations were carried out using the Talys computer code (version number 1.8) \cite{Talys} which is used for the analysis and prediction of nuclear reactions. It is compatible for the simulation of nuclear reactions that involve neutrons, photons, protons, deuterons, tritons, $^{3}$He- and $\alpha$-particles in the 1 keV - 200 MeV energy range, and target nuclides of mass 12 and heavier. 
The cross sections of the $^{112}$Sn($\alpha$,$\gamma$)$^{116}$Te reaction were calculated for combination of different optical model potentials (OMP), level density models (LDM) and strength function models (SFM) in order to investigate the effect of the different nuclear input parameters.

\subsection{Optical Model Potentials}
\label{OMP}

\begin{figure}
\includegraphics {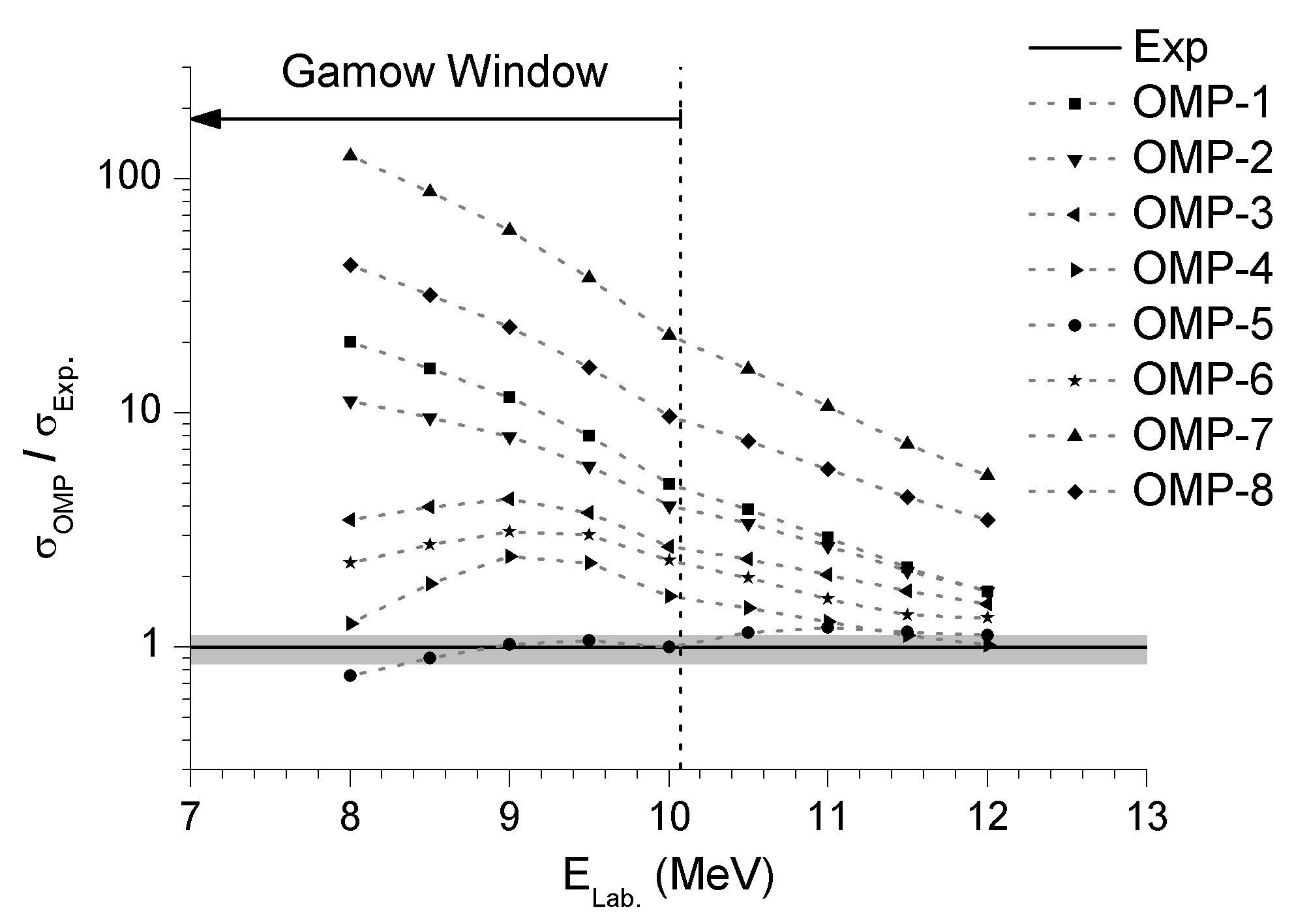}
\caption{The ratio of the calculated cross sections of the different optical model potential (OMP) to the experimental results of the \"{O}zkan et. al. \cite{Ozkan07}. The level density model and strength function were set the constant temperature + Fermi gas model \cite{Gilbert65} (LDM-1) and the Brink-Axel Lorentzian model \cite{Brink57,Axel62} (SFM-2), respectively. Dotted line connecting the points is a guide for the eye.}
\label{fig-1:XS_Ratio}
\end{figure}

The cross sections of the $^{112}$Sn($\alpha$,$\gamma$)$^{116}$Te reaction were calculated for eight different global alpha optical model potential: normal alpha potential \cite{Watanabe58}, McFadden and Satchler \cite{McFadden66}, Demetriou et al. \cite{Demetriou02} (in three version; table 1, table 2, and dispersive model), Avrigeanu et al. \cite{Avrigeanu14}, Nolte et al. \cite{Nolte87}, Avrigeanu et al. \cite{Avrigeanu94}. The level density and strength function model were set to constant temperature-Fermi gas model (LDM-1) and Brink-Axel Lorentzian model (SFM-2), respectively, which are the default settings of the Talys code. The calculated cross sections with different optical model potentials were scaled to experimental cross sections of the \"{O}zkan et. al. \cite{Ozkan07} which was measured most precise and in a wide energy range among the other experiments\cite{Ozkan02, Rapp08, Netterdon15}. The optical model potentials used in the cross section calculation are given at Table \ref{tab-1:OMP} (also labeled in Figure \ref{fig-1:XS_Ratio}). As shown in Figure \ref{fig-1:XS_Ratio}, the best energy dependence of the calculated cross section was the dispersive model of Demetriou et al. \cite{Demetriou02} (OMP-5). This optical potential model almost reproduces the experimental data within the experimental uncertainty except the lowest energy point. All other optical model potentials were significantly higher than the experimental cross sections, especially at the astrophysically relevant energies. 

\begin{table}
\caption{The different optical model potentials which are available in the Talys code. The default option is the normal alpha potential (OMP-1).}
\begin{tabular}{@{}l  l@{}} 

\toprule
{Model no}   &  {Optical model potential}                  \\ 
\midrule
OMP-1 	& Normal alpha potential (1958) \cite{Watanabe58}            \\
OMP-2  	& McFadden and Satchler (1966) \cite{McFadden66}          \\
OMP-3 	& Demetriou et al. (2002) (table 1) \cite{Demetriou02}    \\
OMP-4 	& Demetriou et al. (2002) (table 2) \cite{Demetriou02}    \\
OMP-5	& Demetriou et al. (2002) (dispersive model) \cite{Demetriou02} \\
OMP-6 	& Avrigeanu et al. (2014) \cite{Avrigeanu14}       \\
OMP-7	& Nolte et al. (1987) \cite{Nolte87}        \\
OMP-8 	& Avrigeanu et al. (1994) \cite{Avrigeanu94}      \\
\bottomrule
\end{tabular} \label{tab-1:OMP}
\end{table}

\subsection{Level Density Models}
\label{LDM}

\begin{figure}

\includegraphics {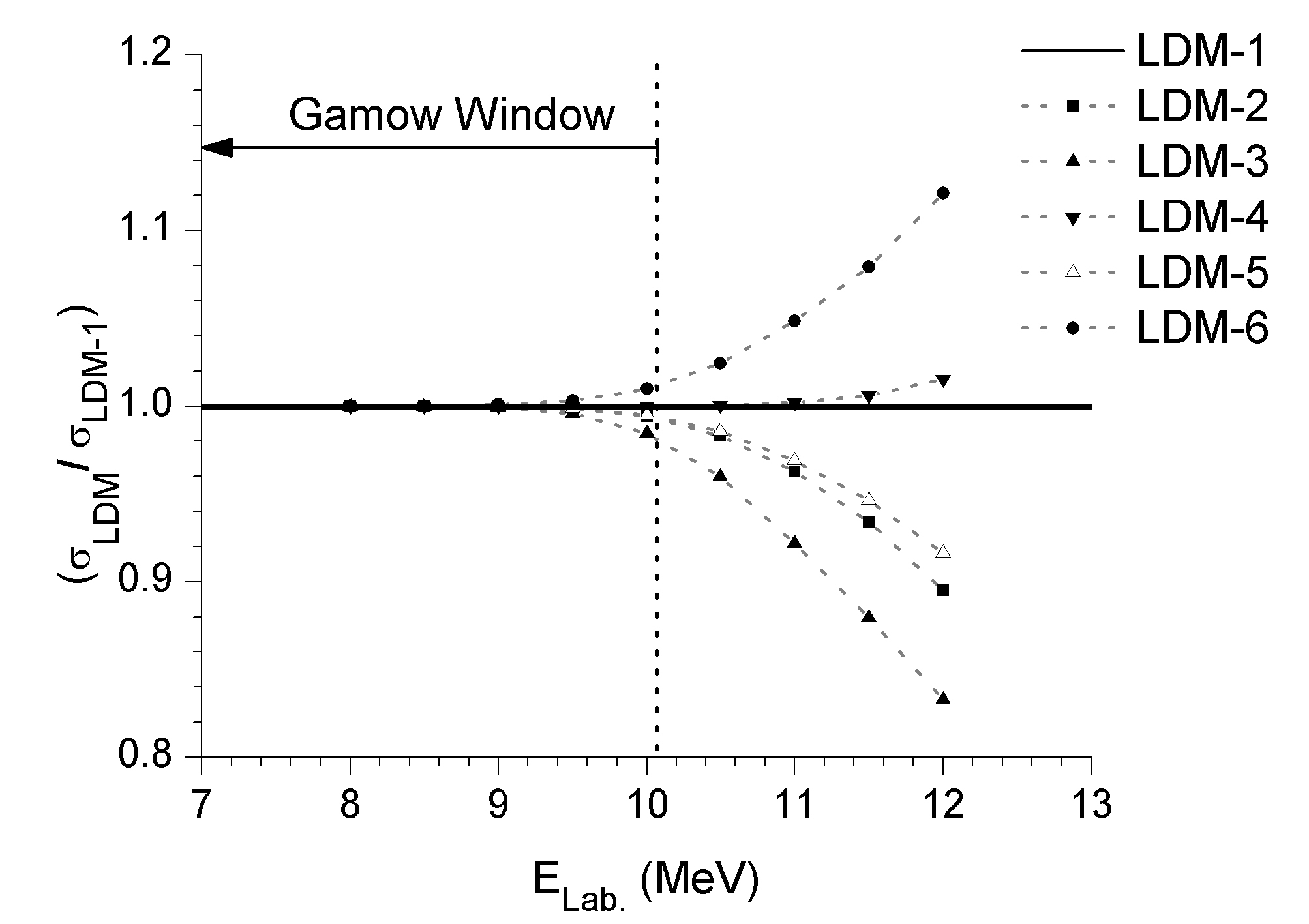}
\caption{The calculated cross sections ratio of the level density model of constant temperature + Fermi gas model \cite{Gilbert65} (LDM-1) to the other level density models (LDM) (see Table \ref{tab-2:LDM}). Dotted line connecting the points is a guide for the eye. }
\label{fig-2:LD_Ratio} 
\end{figure}

The optical model potential and strength function model were set to the dispersive model of Demetriou et al. (OMP-5) and the Brink-Axel Lorentzian model (SFM-2), respectively, in order to investigate the effect of different level density models on the cross section calculation of the $^{112}$Sn($\alpha$,$\gamma$)$^{116}$Te reaction. The cross sections were then calculated with different phenomenological and microscopic level density models which are given in the Table \ref{tab-2:LDM}. Figure \ref{fig-2:LD_Ratio} shows the ratio of the cross section calculation results with different level density models to that with the default level density model (LDM-1). As shown in Figure \ref{fig-2:LD_Ratio}, the cross section deviations are less than 2\% in the Gamow window. With the increasing energy, cross section results increase for Microscopic level densities of \cite{Goriely08}  and \cite{Hilaire13} (LDM-4 and LDM-6) whereas they decrease for Back-shifted Fermi gas model \cite{Dilg01, Demetriou01}, Generalized superfluid model   \cite{Ignatyuk79, Ignatyuk93} and Microscopic level densities of \cite{Hilaire06} (LDM-2, LDM-3 and LDM-5).    

\begin{table}
\caption{Different Level density model which are available in the Talys code. The default option is constant temperature + Fermi gas model (LDM-1). }
\begin{tabular}{@{}l l@{}} 
\toprule

{Model no}   &  {Level density model}                  \\ 
\midrule

LDM-1 	& Constant temperature + Fermi gas model   \cite{Gilbert65}       \\
LDM-2  	& Back-shifted Fermi gas model   \cite{Dilg01, Demetriou01}     \\
LDM-3 	& Generalised superfluid model   \cite{Ignatyuk79, Ignatyuk93}   \\
LDM-4 	& Microscopic level densities (Skyrme force) \cite{Goriely08} \\
        & from Goriely's tables       \\
LDM-5		& Microscopic level densities (Skyrme force) \cite{Hilaire06} \\
        & from Hilaire's combinatorial tables  \\
LDM-6 	& Microscopic LD (temperature dependent HFB, Gogny force) \\
        & from Hilaire's combinatorial tables (2014)   \cite{Hilaire13}      \\
\bottomrule
\end{tabular} \label{tab-2:LDM}
\end{table}

\subsection{Strength Function Models}
\label{SFM}

\begin{figure}

\includegraphics {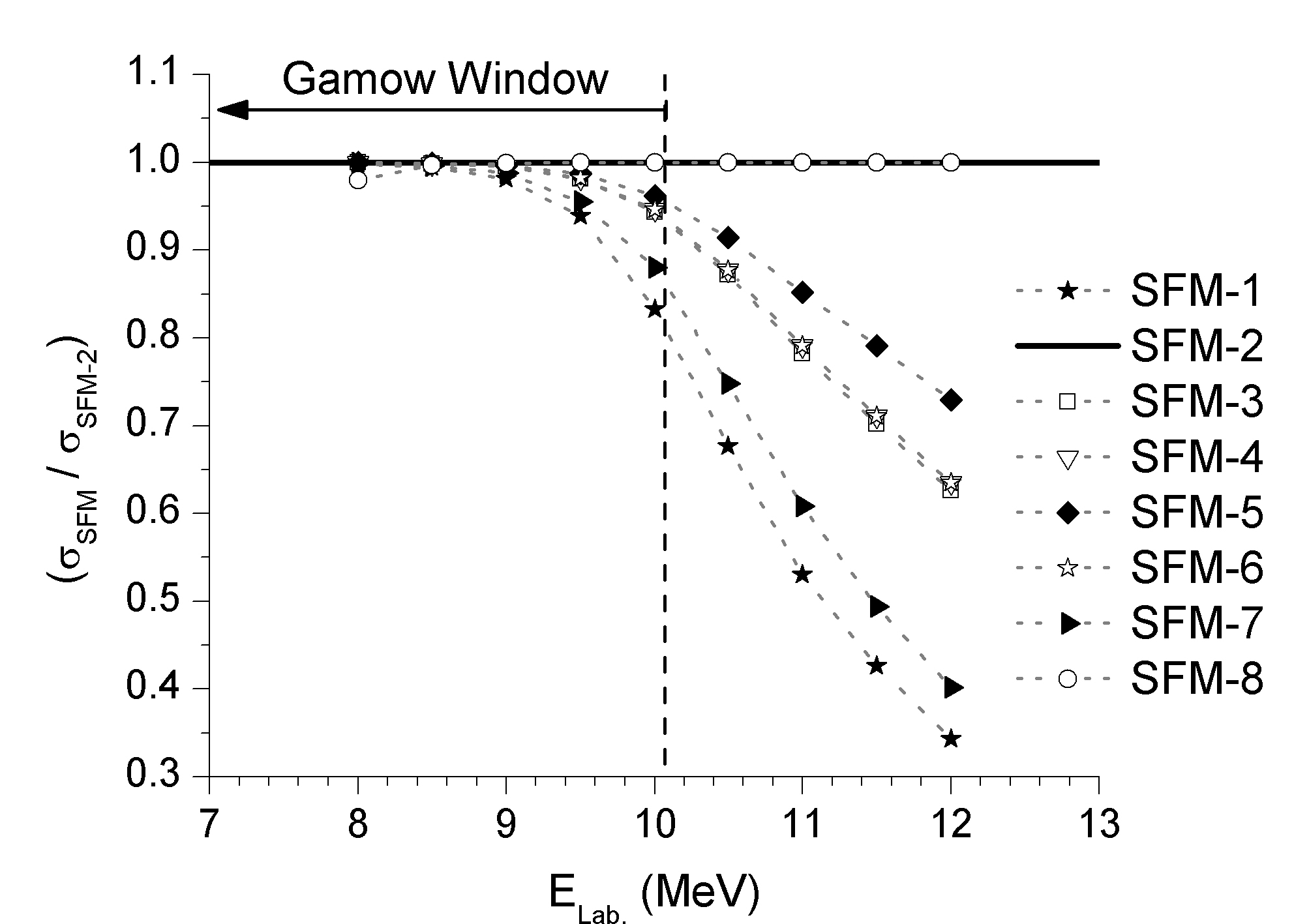} 
\caption{The cross section ratios of different strength function models to the Brink-Axel Lorentzian model \cite{Brink57, Axel62} (see Table \ref{tab-3:SFM}). The optical model potential and level density model were set to dispersive model of Demetriou et al. \cite{Demetriou02} (OMP-5) and Back-shifted Fermi gas model \cite{Dilg01, Demetriou01} (LDM-2), respectively. Dotted line connecting the points is a guide for the eye.}
\label{fig:XS_Ratio_SFM}
\end{figure}

The cross section of the $^{112}$Sn($\alpha$,$\gamma$)$^{116}$Te reaction also depends on the gamma strength function. For this reason, the contributions of eight gamma strength function models to the cross section were investigated which are given in Table \ref{tab-3:SFM}. The optical model potential and level density model were set to the dispersive model of Demetriou et al.\cite{Demetriou02} (OMP-5) and Back-shifted Fermi gas model \cite{Dilg01, Demetriou01} (LDM-2), respectively. Figure \ref{fig:XS_Ratio_SFM} shows the ratio of the cross sections obtained from each strength function model to the cross section from Brink-Axel Lorentzian model (SFM-2). The Brink-Axel Lorentzian (SFM-2) and Gogny D1M HFB+QRPA (SFM-8) models give almost the same results at all energy points, while the other strength function models estimate lower cross section values. The highest difference in the cross section is around 15 percent in the Gamow Window.     

\begin{table}
\caption{Different gamma-ray strength function model which are available in the Talys code. The default option is Brink-Axel Lorentzian model (SFM-2).}

\begin{tabular}{@{}l l@{}} 
\toprule

{Model no}   &  {Strength function model}                  \\ 

\midrule

SFM-1 	& Kopecky-Uhl generalized Lorentzian \cite{Kopecky90,Kopecky93}  \\
SFM-2  	& Brink-Axel Lorentzian \cite{Brink57,Axel62}       \\
SFM-3 	& Hartree-Fock BCS tables \cite{Goriely02}    \\
SFM-4 	& Hartree-Fock-Bogolyubov tables  \cite{Goriely04}    \\
SFM-5		& Goriely's hybrid model \cite{Goriely98} \\
SFM-6 	& Goriely T-dependent HFB   \cite{Hilaire13}   \\
SFM-7		& T-dependent RMF     \cite{Arteaga08}      \\
SFM-8 	& Gogny D1M HFB+QRPA   \cite{Martini2014}   \\

\bottomrule

\end{tabular} \label{tab-3:SFM}
\end{table}

\section{Discussion and Conclusion}
\label{RaD}

Based on the results of cross section calculations with different optical model potentials, level density models and strength function models, it is found that cross section of the $^{112}$Sn($\alpha$,$\gamma$)$^{116}$Te reaction has a strong dependence on the optical model potentials. On the other hand, cross section calculations with different level density models and strength function models give comparable results in the Gamow window. As a result, the cross section calculation with the combination of dispersive model of Demetriou et al. \cite{Demetriou02} (OMP-5), Back-shifted Fermi gas level density model (LDM-2) and the Brink-Axel Lorentzian strength function model (SFM-2) best reproduced the experimental cross sections.

Because the charged-particle cross section is highly energy dependent, extrapolation of the cross section to the lower energies and the comparison between theoretical and experimental results in the low energy region are very difficult. The astrophysical $S$ factor removes the part of the strong energy dependence of the cross section by accounting for the s-wave Coulomb barrier transmission $\exp (-2\pi \eta)$ at low energies. For this reason, it is a useful tool for the analysis of charged-particle reactions. The $S$ factor is defined as \cite{Rolfs87}

\begin{equation}
 S(E)=\sigma (E) E e^{2\pi \eta} \quad.
\end{equation}

\begin{figure}
\includegraphics{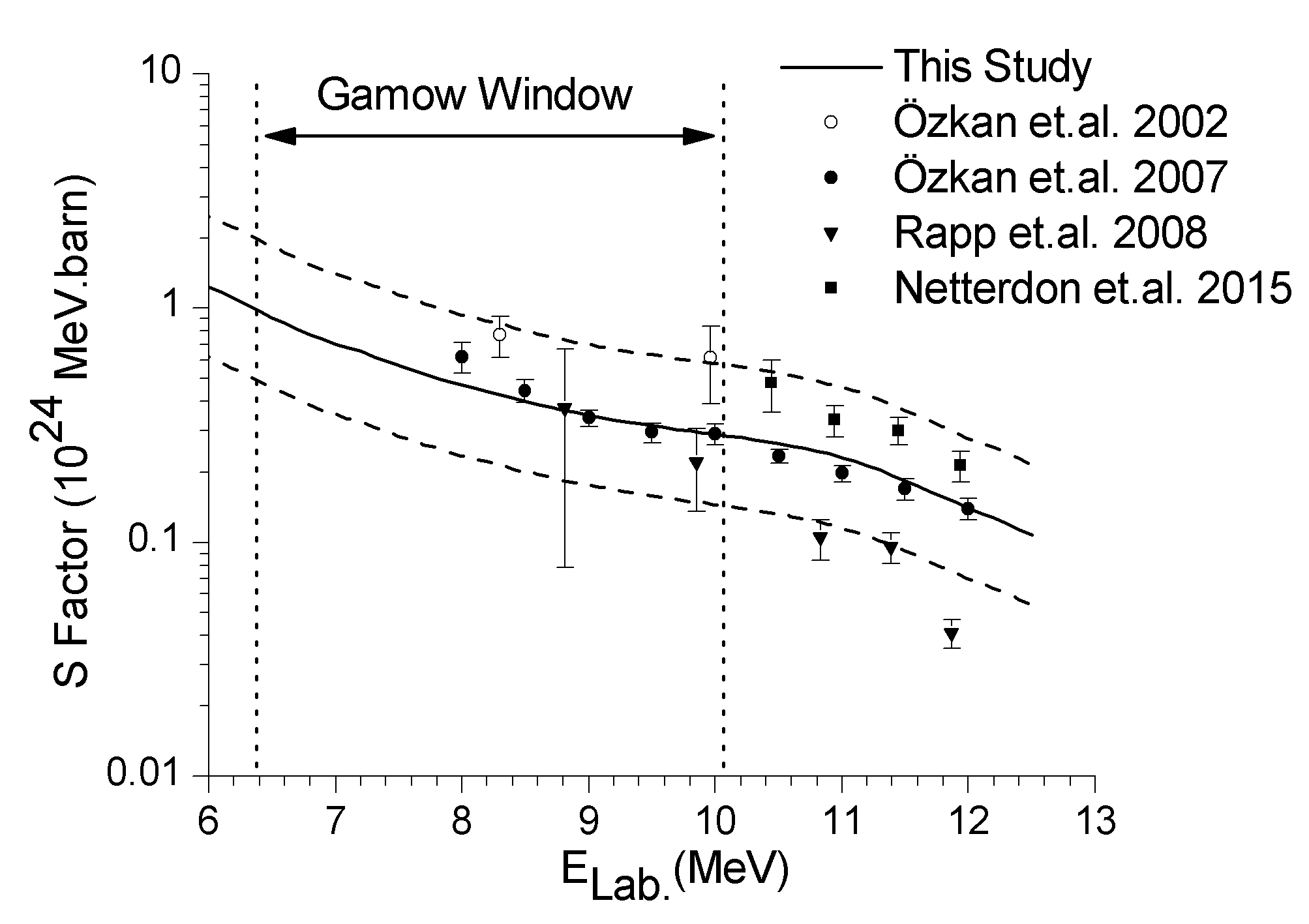} 
\caption{Theoretically and experimentally calculated astrophysical $S$ factors. Gamow window is also shown in the figure for the temperature of 3 GK. Dashed lines are showing the calculated $S$ factors multiplied by 0.5 and 2.}
\label{fig:SF_comparison}
\end{figure}

Where $\eta$ is the Sommerfeld parameter, as defined in reference \cite{Rolfs87}. The astrophysical $S$ factors were calculated from the cross sections with the best model combination (OMP-5, LDM-2, SFM-2) and compared with experimental results. The experimental results of \"{O}zkan et. al. \cite{Ozkan07} were well described by the theoretical calculation with this model combination (see Figure \ref{fig:SF_comparison}).

The reaction rates which are needed for the $p$ process simulation were also calculated using the best model combination (OMP-5, LDM-2, SFM-2). The average reaction rate per particle pair at a given stellar temperature $T^{*}$ is defined by:

\begin{equation}
\begin{aligned}
\left\langle \sigma v \right\rangle ^{*} = & \left(\frac{8}{\pi\mu}\right)^{1/2} \frac{1}{\left(kT^{*}\right)^{3/2}} \\ 
& \int^{\infty}_{0} \sigma^{*}_{(\alpha,\gamma)} (E) E \exp\left(- \frac{E}{kT^{*}} \right)dE , 
\end{aligned}
\end{equation}

by folding the stellar reaction cross section $\sigma^{*}_{(\alpha,\gamma)}(E)$ with the Maxwell-Boltzmann velocity distribution of the nuclei \cite{Rolfs87}. Here $\mu$ is the reduced mass of the system. The nuclei can also be found in excited states in the stellar plasma, therefore the stellar reaction cross section  $\sigma^{*}=\sum_{\lambda\nu}\sigma^{\lambda\nu}$  includes transitions from all populated target states $\lambda$ to all energetically possible final states $\nu$, whereas a laboratory cross section $\sigma^{lab}=\sum_{\nu}\sigma^{0\nu}$ only accounts for transitions from the ground states of the target. The ratio of the stellar to laboratory reaction cross section  $\sigma^{*}$/$\sigma^{lab}$ is called stellar enhancement factor. Since there is no low-lying excited state in $^{112}$Sn, stellar enhancement factor of the $^{112}$Sn($\alpha$,$\gamma$)$^{116}$Te reaction is negligible at the $p$ process temperature of $2.0 \leq T_{9} \leq3.0$ (where $T_{9}$ is the temperature in GK). 

\begin{figure}

\includegraphics{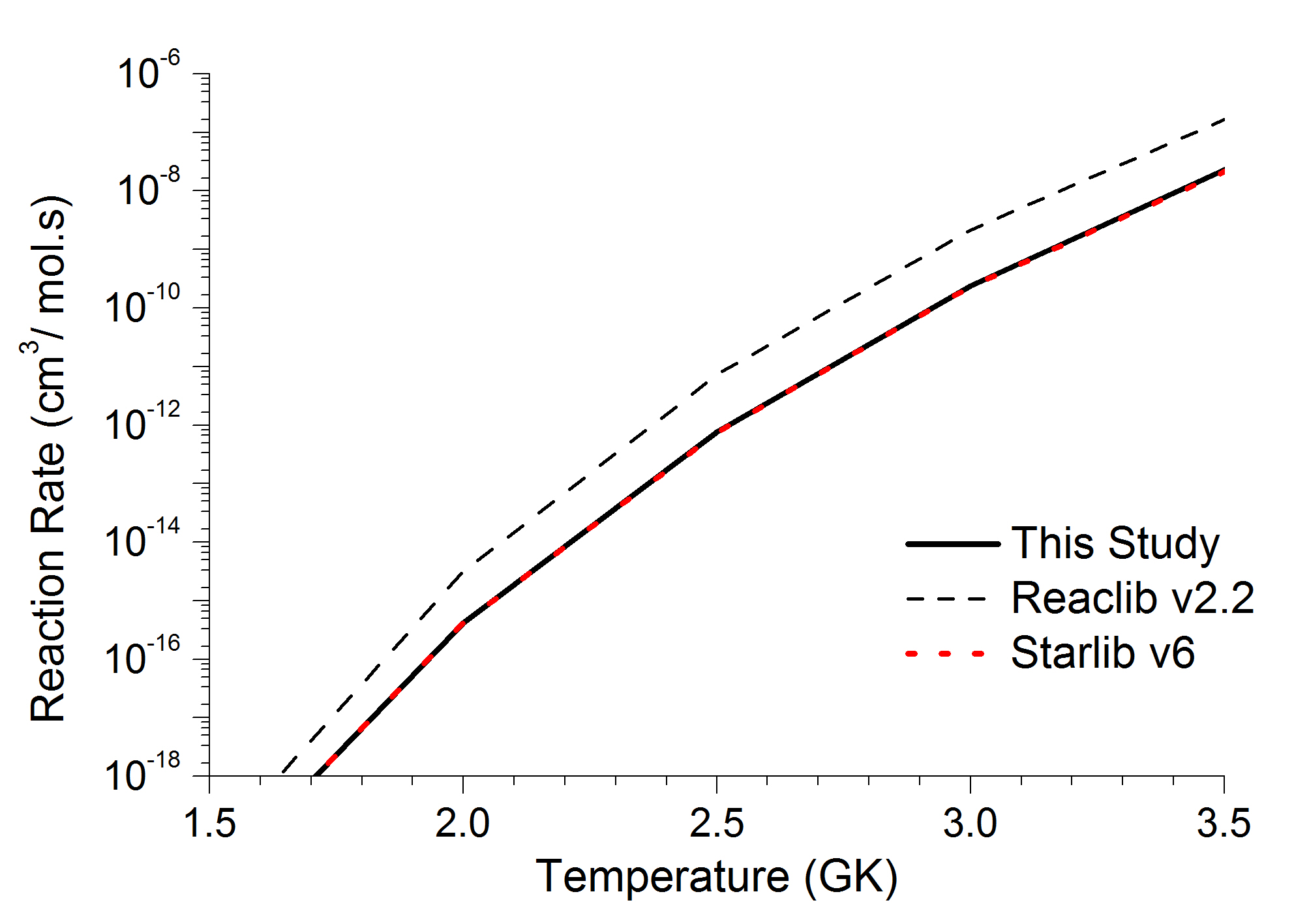}
\caption{The comparison of the calculated reaction rates with the Reaclib $v2.2$ \cite{Reaclib} and Starlib $v6$ \cite{Starlib} rate libraries for the $^{112}$Sn($\alpha$,$\gamma$)$^{116}$Te reaction. The reaction rates are shown for the p process temperatures, for all temperature scale see Table \ref{tab-4:rates}.}
\label{fig:rates}
\end{figure}

Reaction rate results are given in Table \ref{tab-4:rates}. Figure \ref{fig:rates} shows the comparison of the calculated reaction rates with those in the Reaclib $v2.2$ (data set; ths8(v4)) \cite{Reaclib} and Starlib $v6$ \cite{Starlib}. It is found that calculated reaction rates are in excellent agreement with those reported by Starlib $v6$ \cite{Starlib} while they are considerably lower than those reported by Reaclib $v2.2$ \cite{Reaclib}.

\begin{table}
\caption{The calculated reaction rates with the best model parameters (OMP-5, LDM-2, SFM-2).
}

\begin{tabular}{@{}c c@{}} \toprule

{Temperature }   &  {Reaction Rate}                  \\ 
{(GK) }   &  {cm$^{3}$/mol.s}                  \\ 
\midrule

0.25	& 3.27E-65   \\ 
0.3		& 2.55E-59   \\ 
0.4		& 1.58E-50   \\ 
0.5		& 2.44E-44   \\ 
0.6		& 1.88E-39   \\ 
0.7		& 1.48E-35   \\ 
0.8		& 1.98E-32   \\ 
0.9		& 7.00E-30   \\ 
1			& 9.32E-28   \\ 
1.5		& 1.31E-20   \\ 
2			& 4.13E-16   \\ 
2.5		& 7.55E-13   \\ 
3			& 2.42E-10   \\ 
3.5		& 2.31E-8   \\ 
4			& 8.22E-7   \\ 
5			& 9.19E-5   \\ 
6			& 0.00154   \\ 
7			& 0.0113   \\ 
8			& 0.0504   \\ 
9			& 0.158   \\ 
10		& 0.374   \\ 

\bottomrule

\end{tabular} 
\label{tab-4:rates}
\end{table}

The theoretical calculations of the cross sections are as important as the experimental efforts to study the nucleosynthesis theory. In this study, the cross sections of the $^{112}$Sn($\alpha$,$\gamma$)$^{116}$Te reaction were calculated with different optical model potentials, level density models and strength function models in order to understand the effect of different nuclear parameters. The conclusions of this study can be summarized as follows: 

- The cross section calculations are very sensitive to global optical model potentials (OMP) for $^{112}$Sn($\alpha$,$\gamma$)$^{116}$Te reaction. The sensitivity to optical model potentials are increasing with decreasing energy. 

- The different level density models contribution to cross section calculations are very limited in the Gamow windows. 

- The cross section difference for different strength function models are less than 15\% in the Gamow window.  

- The optical model potential of dispersive model by Demetriou et al. with the combination of Back-shifted Fermi gas level density model and the Brink-Axel Lorentzian strength function model best reproduces the experimental $S$ factor (or cross section) results by \"{O}zkan et. al. \cite{Ozkan07}.  

- The calculated $S$ factors agree with all of the experimental results within a factor of 2, except the highest energy point by Rapp et al. \cite{Rapp08}. 

- The reaction rate library by Reaclib $v2.2$ \cite{Reaclib} overestimated the reaction rates. Calculated reaction rate results are 7-10 times lower than those by Reaclib $v2.2$ in the $p$ process temperature of 2-3 GK. 

-The calculated reaction rates are in excellent agreement with Starlib $v6$ \cite{Starlib} in the temperature range of 1-3.5 GK. 

As a result, new reaction rate values (see Table \ref{tab-4:rates}) for $^{112}$Sn($\alpha$,$\gamma$)$^{116}$Te reaction are suggested to the $p$ process nucleosynthesis simulation. The investigation of other nuclear reactions related to $p$ process nucleosynthesis will help to develop a reliable nucleosynthesis theory.

\end{document}